\documentclass[referee]{raa}           
\usepackage{graphicx,times}
\usepackage{natbib}
\usepackage{amssymb,amsmath}
\bibpunct{(}{)}{;}{a}{}{,}

\usepackage{supertabular}
 \usepackage{threeparttable}
 \usepackage{longtable}
\usepackage{color}
 \usepackage{multirow}
 \usepackage{url}
 \usepackage{txfonts}

\usepackage[a4paper=true,pagebackref=true]{hyperref}
\hypersetup{pdftitle = The title of my PDF, pdfauthor = My name, pdfsubject= The subject, pdfkeywords = keyword1 keyword2 keyword3} 
\hypersetup{colorlinks = true, linkcolor = green, anchorcolor = red, citecolor = blue, filecolor = red, pagecolor = red, urlcolor = red}

\begin{document}

  \title{Constraints on individual supermassive binary black holes using observations of PSR~J1909$-$3744 
}

 \volnopage{ {\bf 2019} Vol.\ {\bf X} No. {\bf XX}, 000--000}
   \setcounter{page}{1}

   \author{Yi Feng\inst{1,2,3}
   \and Di Li \inst{1,3} 
   \and Yan-Rong Li \inst{4}
   \and Jian-Min Wang \inst{4}
   }

   \institute{ National Astronomical Observatories, Chinese Academy of Sciences, Beijing 100012, China ; {\it yifeng@nao.cas.cn}\\
        \and
           University of Chinese Academy of Sciences, Beijing 100049, China \\
	\and
CAS Key Laboratory of FAST, National Astronomical Observatories, Chinese Academy of  Sciences, Beijing\\
\and
Institute of High Energy Physics, Chinese Academy of Sciences,19B Yuquan Road, Beijing 100049, China\\
\vs \no
}

\abstract{ 
We perform a search for gravitational waves (GWs) from several supermassive binary black hole (SMBBH) candidates (NGC 5548, Mrk 231, OJ 287, PG 1302-102, NGC 4151, Ark 120 and 3C 66B) in long-term timing observations of the pulsar PSR~J1909$-$3744 obtained using the Parkes radio telescope. No statistically significant signals were found. We constrain the chirp masses of those SMBBH candidates and find the chirp mass of NGC 5548 and 3C 66B to be less than  $2.4 \times 10^9\,\rm M_{\odot}$ and $2.5 \times 10^9\,\rm M_{\odot}$ (with 95\% confidence), respectively. Our upper limits remain a factor of 3 to 370 above the likely chirp masses for these candidates as estimated from other approaches. The observations processed here provide upper limits on the GW strain amplitude that improve upon the results from the first Parkes Pulsar Timing Array data release by a factor of 2 to 7. We investigate how information about the orbital parameters can help improve the search sensitivity for individual SMBBH systems. Finally, we show that these limits are insensitive to uncertainties in the Solar System ephemeris model.
\keywords{black hole physics --- gravitational waves --- pulsars: general ---
pulsars: individual (PSR~J1909$-$3744) 
}
}

   \authorrunning{Y. Feng et al. }            
   \titlerunning{Constraints on individual supermassive binary black holes using observations of PSR~J1909$-$3744}  
   \maketitle

%
\section{Introduction} \label{sec:intro}
Pulsar timing arrays (PTAs) can be used to detect and study gravitational waves (GWs) in the nanohertz frequency band \citep{Detweiler1979,Foster_Backer90}. The primary source of GWs in this band are thought to be inspiralling supermassive binary black holes (SMBBHs), formed in the aftermath of galaxy mergers \citep{1980Natur.287..307B}.
The separations of these binary systems are generally at sub-parsec scales, which are usually too small to be resolved with current electromagnetic instruments. This makes identification of SMBBHs and measurements of their physical properties (such as the black hole mass and orbital parameters) particularly challenging. 
Thus far there have been a number of SMBBH candidates identified through several lines of approaches,
although the nature of these candidates are still under investigation (e.g., \citealt{1997ApJ...484..180S}, \citealt{2012ApJ...759..118B},\citealt{2015Natur.518...74G}, \citealt{2015ApJ...809..117Y}, \citealt{2016ApJ...822....4L}). 

There has recently been significant progress in searches for GWs from individual SMBBHs in PTA data sets (\citealt {2010MNRAS.407..669Y},
\citealt {2014MNRAS.444.3709Z},
\citealt {2016MNRAS.455.1665B},
\citealt {NANOGrav11yrCW}).
Although such projects have not detected GWs yet, the first detection is likely in the next one or two decades.  For instance,
\cite{2017NatAs...1..886M} claimed that GWs from at least one nearby SMBBH will be detected within 10 years if the GW background (GWB) can be successfully isolated. Once detected, the SMBBH can also be followed up by electromagnetic observations, thus enabling a multi-messenger view of the black hole system \citep{2013CQGra..30v4013B}. Furthermore, the detection of SMBBH systems can yield direct information about the masses and spins of the black holes \citep{2012PhRvL.109h1104M}, which would shed light on the formation and evolution of supermassive black holes. 
The precisely measured distance of PSR~J0437$-$4715 by \citet{2016MNRAS.455.1751R} would allow the
polarization angle of GWs generated by OJ 287 to be measured to better than $8^{\circ}$ \citep{2018MNRAS.481.2249C}.
Chen and Zhang (2018) also showed that measurement uncertainties of the spin and quadrupole scalar from future PTAs are comparable or even smaller than those in the optical measurements.

\cite{2001ApJ...562..297L} using pulsar observations to place
upper limits on the mass ratio of hypothetical SMBBHs in the \cite{1998AJ....115.2285M} sample of galaxies that were as small as 0.06 for orbital periods of $\sim$ 2000 days.
This work was continued and updated by \cite{ma2016}, which constrained the mass ratios of hypothetical SMBBHs in nearby galaxies using PTA limits \citep{2014MNRAS.444.3709Z,2016MNRAS.455.1665B}. \cite{sudou2003} detected apparent periodic variations in the radio core position of the nearby elliptical galaxy 3C 66B with a period of 1.05\,yr and suggested the presence of a SMBBH with a total mass of $5.4 \times 10^{10}\,\rm M_{\odot}$ and a mass ratio of 0.1.
\cite{jenet2004} subsequently placed limits on the mass and eccentricity of the proposed SMBBH and ruled out the adopted system with 95\% confidence. 

In this work, we use data from long-term timing observations of the pulsar, PSR~J1909$-$3744, from the Parkes Pulsar Timing Array project to place limits
on the chirp mass of a few well-known SMBBH candidates using a method similar to that in \cite{jenet2004}. Our chosen SMBBH candidates, listed below, have all been monitored in detail and show evidence for a SMBBH system: 
\begin{itemize}
\item The well-known BL Lac object OJ 287 exhibits 
a $\sim$12-year periodicity in the 100-year long optical light curve (\citealt{1997ApJ...484..180S}).
\item NCG 4151 (\citealt{2012ApJ...759..118B}) exhibits periodic variations in the H$\alpha$ light and radial velocity curves. 
\item NGC 5548 (\citealt{2016ApJ...822....4L}) shows evidence for periodicites in both the long-term continuum light curves and emission line profile variations.
\item PG 1302$-$102 exhibits periodicity in optical light curve, which is well modeled by the relativistic Doppler boosting of the secondary mini-disk \citep{2015Natur.518...74G}\footnote{There are several modern time-domain surveys that have released candidate samples for SMBBHs through periodicity searches (\citealt{Graham2015, Charisi2016, Liu2016}). Most of these candidates are fairly distant so that their GW emissions are clearly too weak to detect with the current PTA datasets. In addiation, due to the limited temporal baselines (generally $\sim$1.5 cycles of the periods), the reported periodicities may suffer high false alarm probabilities  (\citealt{Vaughan2016}). We thereby only select PG~1302$-$102 reported by \cite{2015Natur.518...74G}, the most significant candidate found in the Catalina Real-Time Transient Survey.}.
\item Ark 120 (\citealt{2019ApJS..241...33L}) exhibits a wave-like pattern in the optical emissions with a total temporal baseline over four decades and the H$\beta$ integrated flux series varies with a similar behaviour.
\item Mrk 231 was found to have a flux deficit at UV wavelength,
which could signify a SMBBH system \citep{2015ApJ...809..117Y}.
\item 3C66B: since \cite{jenet2004} ruled out the original model of \cite{sudou2003}, \cite{Iguchi2010} performed follow-up observations and still found evidence of a SMBBH system, but with significantly lower SMBBH mass estimates than were previously assumed. Such low masses imply that the expected GW emission would be below the detection thresholds of previous all-sky searches using PTA data \citep[see, e.g.,][]{2014MNRAS.444.3709Z}.
\end{itemize}

Table~\ref{tab1} also lists the basic properties of the above candidates. Some of the SMBBH candidates mentioned above have published estimates of the full orbital parameters for the postulated SMBBH systems (e.g., NGC 5548, 3C 66B and OJ 287). Many (but not all) of the previous PTA studies usually did not use such information when placing upper limits on the chirp mass of such systems. Here, we investigate in detail how the extra information, e.g., the eccentricity of the orbit and the inclination angle of the binary orbit with respect to the line of sight, can help improve the search sensitivity. 
Recently, it has been shown that the GWB constraints are sensitive to the choice of the Solar System ephemeris (SSE) adopted in the data analysis \citep{2018ApJ...859...47A}. We explore how the choice of the SSE influences the analysis results presented here. 

The organization of this paper is as follows. Section~\ref{sec:signal} describes the expected signature of GW emission from a general binary system. The observations of PSR~J1909$-$3744
are described in Section~\ref{sec:obs}. 
Section~\ref{sec:result} presents the limit on the chirp mass for our SMBBH candidates. We also discuss here how further information about the orbital parameters (if they were available)  would influence the limit and how the limit changes with those choice of SSE model used. We conclude in Section~\ref{sec:discussion}.

\section{THE SIGNATURE OF A SMBBH}
\label{sec:signal}
We adopt a quadrupole formalism for estimating the signature of GW signals in pulse arrival times (ToAs) from pulsar observations. The pulsar timing residuals induced by a single GW source can be written as
\begin{equation}
\begin{split}
s(t, \hat{\Omega}) = &(F^+(\hat{\Omega})\cos2\psi + F^{\times}(\hat{\Omega})\sin2\psi) {\bigtriangleup}s_{+}(t)\\
&+ (F^+(\hat{\Omega})\sin2\psi - F^{\times}(\hat{\Omega})\cos2\psi)  {\bigtriangleup}s_{\times}(t),
\end{split}
\end{equation}
where $\hat{\Omega}$ is a unit vector defining the direction of GW propagation, $\psi$ is the GW polarization angle. The two functions $F^+$ and $F^{\times}$ are the geometric factors which only depend on the GW source position relative to a given pulsar \citep{2011MNRAS.414.3251L}
\begin{equation}
\begin{split}
F^{+} = &\frac{1}{4(1-\cos \theta)}\{(1+\sin^2\delta)\cos^2\delta_p\cos[2(\alpha-\alpha_p)] \\
&-\sin 2\delta\sin2\delta_p\cos(\alpha-\alpha_p)+\cos^2\delta(2-3\cos^2\delta_p)\},\\
F^{\times} = &\frac{1}{2(1-\cos \theta)}\{\cos\delta\sin2\delta_p\sin(\alpha-\alpha_p) \\
&-\sin\delta\cos^2\delta_p\sin[2(\alpha-\alpha_p)] \},
\end{split}
\end{equation}
where $\cos\theta = \cos\delta\cos\delta_p\cos(\alpha-\alpha_p)+\sin\delta\sin\delta_p$ with $\theta$
being the angle between the GW source and pulsar
direction with respect to the observer, $\delta (\delta_p)$ and $\alpha (\alpha_p)$ are
the declination and right ascension of the GW source (pulsar),
respectively.

The GW-induced pulsar timing residuals can be expressed as the combination of two terms - the Earth term $s_{+,\times}(t)$ and the pulsar term $s_{+,\times}(t_p)$:
\begin{equation}
\bigtriangleup s_{+,\times}(t) = s_{+,\times}(t) - s_{+,\times}(t_p),
\end{equation}
\begin{equation}
t_p = t - d_p(1-\cos\theta)/c,
\end{equation}
where $d_p$ is the pulsar distance. $s_+(t)$ and $s_{\times}(t)$ are source-dependent functions,
and at Newtonian order take the same forms for GWs emitted
by SMBBHs in eccentric orbits as \citep{2016ApJ...817...70T}:
\begin{equation}
\begin{split}
s_+(t) = &\sum \limits_{n} - (1 + \cos^2\iota)[a_n \cos(2\gamma) - b_n \sin(2\gamma)]\\ 
&+ (1 - \cos^2 \iota)c_n,\\
s_{\times}(t) = &\sum \limits_{n}2\cos\iota[b_n \cos(2\gamma) + a_n \sin(2\gamma)],
\end{split}
\end{equation}
where 
\begin{equation}
\begin{split}
a_n =&- \zeta \omega^{-1/3}[J_{n-2}(ne) - 2eJ_{n-1}(ne) + (2/n)J_{n}(ne)\\ 
&+ 2eJ_{n+1}(ne) - J_{n+2}(ne)\sin[nl(t)],\\
b_n =&\ \zeta \omega^{-1/3}\sqrt{1 - e^2}[J_{n-2}(ne) - 2J_{n}(ne)\\ 
&+ J_{n+2}(ne)]\cos[nl(t)],\\
c_n =&\ (2/n)\zeta \omega^{-1/3}J_{n}(ne)\sin[nl(t)].
\end{split}
\end{equation}
Here, $\zeta = (GM)^{5/3}/c^4D_L$ is the amplitude parameter, where
$M$ is the binary chirp mass defined as $M^{5/3} = m1 m2 (m1 + m2)^{-1/3}$ with $m1$
and $m2$ being the binary component masses,
$D_L$ is the luminosity distance of the binary, $e$ is the
eccentricity, and $\omega = 2\pi f$. $f$ is the orbital frequency;
$l(t)$ is the mean anomaly; $\gamma$ is the initial angle of periastron; and $\iota$ is the inclination angle of the binary orbit with respect
to the line of sight. In the following,
$f$ and $M$ refer to the observed redshifted values, such that
$f_r = f(1+z)$ and $M_r = M/(1+z)$, where $f_r$ and $M_r$ are
rest frame values, and $z$ is the cosmological redshift of the
binary.
We assume no binary evolution over typical pulsar timing baselines. The Earth term $s_{+,\times}(t)$ and pulsar term $s_{+,\times}(t_p)$ are computed by assuming that the binary's mean orbital frequency $f$ and eccentricity $e$ remain constant over the total time span of our observations of a given pulsar.
Meanwhile, we assume binary evolution over the lag time between the Earth and pulsar term signals. 
$e$ and $f$ for the pulsar term are obtained by solving the coupled differential equations \citep{1964PhRv..136.1224P}
\begin{align} \label{eq:pulsar}
\frac{{\rm d}f}{{\rm d}t} &= \frac{48}{5\pi}(\frac{GM}{c^3})^{5/3}(2\pi  f)^{11/3}\frac{1+\frac{73}{24}e^2+\frac{37}{96}e^4}{(1-e^2)^{7/2}},\nonumber\\
\frac{{\rm d}e}{{\rm d}t} &= -\frac{304}{15}(\frac{GM}{c^3})^{5/3}(2\pi f)^{8/3}e\frac{1+\frac{121}{304} e^2}{(1-e^2)^{5/2}}.
\end{align}

\section{TIMING OBSERVATIONS}
\label{sec:obs}
We use data from long-term timing observations of the pulsar PSR~J1909$-$3744, which is one of the best high-time-precision pulsars. We have processed two, related data sets. The first (known as ``PPTA DR1") is the initial PPTA data release
\citep{2013PASA...30...17M} and we make use of the data collection for this data set provided by \cite{2016MNRAS.455.3662M}.  We also process the more up-to-date data set that was used to constrain ultralight scalar-field dark matter by \cite{2018PhRvD..98j2002P} (hereafter we call this the ``P18" data set).

The DR1 data set was acquired between March 2005 and March 2011 for 20 pulsars.
The P18 data set consists of observations for 26 pulsars collected
between February 5, 2004 and January 31, 2016. It includes the
DR1 data along with some earlier data for some pulsars.
The observing cadence is normally once every two to
three weeks. At most observing epochs, each pulsar was observed
in three radio bands (10, 20 and 50\,cm) with a typical integration
time of one hour.

Pulsar times of arrival were fitted with a timing model using the standard
TEMPO2 \citep{hobbs2006} software package. Typical parameters
included in this fit are the five astrometric parameters (sky position, proper motion, and parallax), spin frequency, spin-down rate, dispersion measure, and (when applicable)
binary orbital parameters. Additionally, constant offsets
or jumps were fitted among ToAs collected with different
receiver/backend systems. 
For the DR1 data set, we used TT(BIPM2011) as the reference time standard and adopted the JPL DE421 solar system ephemeris (SSE) model.
For the more up-to-date data set, we used TT(BIPM2015) and adopted the JPL DE418 SSE model.

The pulsars in the data sets are not equally sensitive to GW signatures.  For the work that we describe here we simply make use of a single pulsar. This significantly simplifies the analysis procedures and provides a baseline result for more detailed studies. We note that selecting a single pulsar dataset can lead to biased bounds (which we discuss later), but this work is primarily a proof of concept and we are currently very unlikely to detect the GW signature from a SMBBH even with the full sensitivity obtained using all pulsars in an array.  Such a detailed analysis will be presented by the PPTA team when the second data release has been completed. We therefore use observations of PSR~J1909$-$3744. Similar to \cite{2014MNRAS.444.3709Z} and \cite{2016MNRAS.455.3662M}, ToAs of the best band (i.e. where the lowest rms timing residuals are seen and 10\,cm observing band for  PSR~J1909$-$3744) have been selected after correcting for dispersion measure variations.

\section{RESULTS AND DISCUSSION}
\label{sec:result}

\begin{figure*}
\includegraphics[width=1\textwidth]{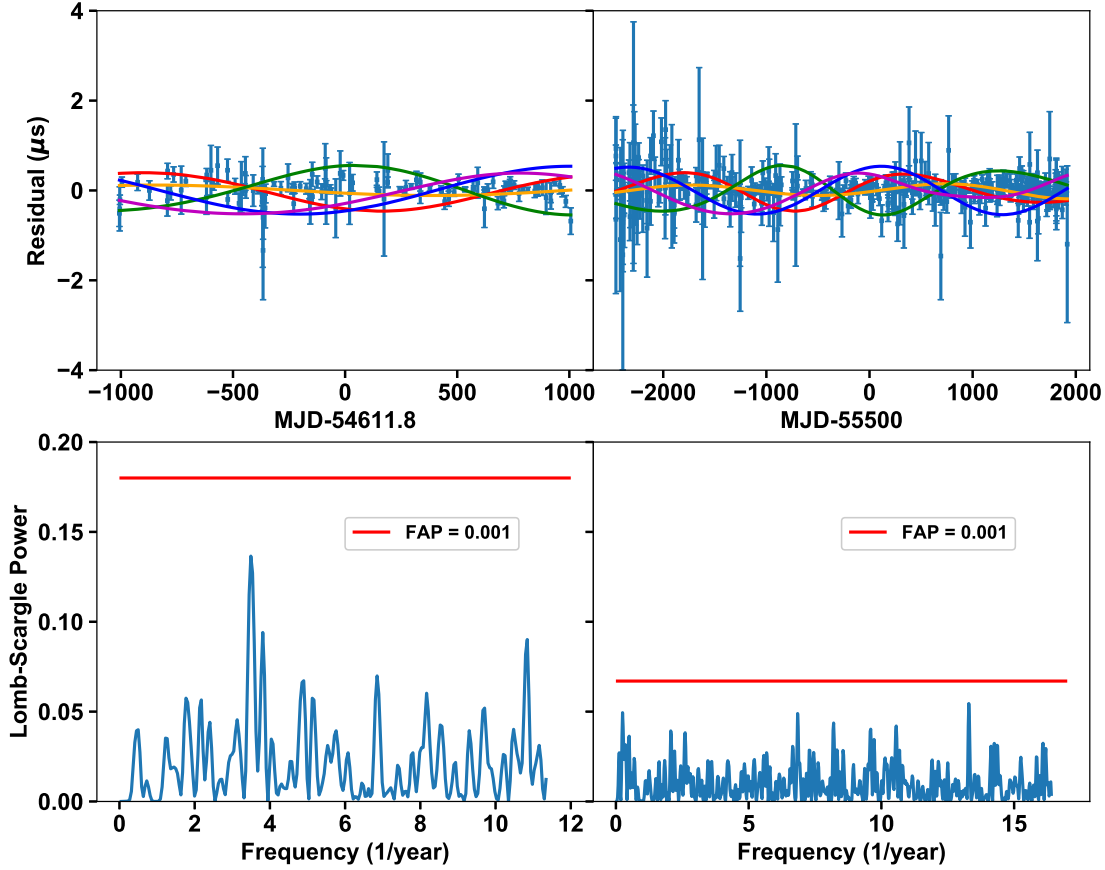}
\caption{Top: Timing residuals (blue dots) for PSR~J1909$-$3744. For each data set, we also show 5 realizations of pre-fit timing residuals (colored lines) induced by NGC 5548 with the pulsar term included. See the text for parameters of these 5 realizations of GW-induced timing residuals. Bottom: The corresponding Lomb-Scargle periodogram.
The horizontal red line corresponds to our detection threshold of $\rm{FAP} = 10^{-3}$. Left: DR1 data set. Right: P18 data set. 
}
\label{fig1}
\end{figure*}

The timing residuals from PSR~J1909$-$3744 were searched
for the signature of GWs, using a Lomb-Scargle
periodogram (LP; \citealt{1976Ap&SS..39..447L}; \citealt{1982ApJ...263..835S}).
There have recently been lots of PTA data analysis methods for single GW sources developed (\citealt {2010MNRAS.407..669Y}, \citealt {2012ApJ...756..175E}, \citealt {2013CQGra..30v4004E}, \citealt {2014MNRAS.444.3709Z}, \citealt {2014ApJ...795...96W}, \citealt {2015MNRAS.449.1650Z}, \citealt {2015ApJ...815..125W}, \citealt {2016MNRAS.461.1317Z}, \citealt {2016ApJ...817...70T}), ranging from frequentist to Bayesian techniques (e.g., \citealt {2013CQGra..30v4004E}, \citealt {2016ApJ...817...70T}), and from Earth-term-only approaches to a coherent inclusion of both Earth terms and pulsar terms (e.g., \citealt {2014ApJ...795...96W}, \citealt {2015ApJ...815..125W},
\citealt {2016MNRAS.461.1317Z}). Although Lomb-Scargle periodogram is sub-optimal, this simplifies the analysis procedures and provides a baseline result for more detailed studies. And we defer analysis using optimal methods including ellipticity and the evolution of binary orbit during the period of observation to a future study.
The timing residuals and Lomb-Scargle power are shown in the top and bottom panel of Figure~\ref{fig1} for the two data sets respectively.
The LP for DR1 data set is shown in the bottom left panel of Figure~\ref {fig1}.
The periodogram power was determined for 250 frequencies
ranging from 1/22.0 $\rm yr^{-1}$ to 11.4 $\rm yr^{-1}$, with a resolution of 1/22.0 $\rm yr^{-1}$. This corresponds to a frequency-oversampling
factor of four as the time span of the DR1 data is 5.5\,yrs. 
In order to determine the statistical significance of a given peak, we generate synthetic data sets of Gaussian deviates with the same sampling as in the real data and find the largest periodogram power for each such data set. Using the distribution of largest periodogram power of $10^5$ synthetic data sets, the periodogram power of a significant peak is 0.18 assuming a False-Alarm probability (FAP) of 0.001 .
Hence there is no significant peak in this LP.

The LP for P18 data set is shown in the bottom right panel of Figure~\ref {fig1}.
The periodogram power was calculated for 786 frequencies
ranging from 1/48.0 $\rm yr^{-1}$ to 16.4 $\rm yr^{-1}$, with a resolution of
1/48.0 $\rm yr^{-1}$. This corresponds to a frequency-oversampling
factor of 4 as the time span of P18 data is 12\,yrs. Similar to PPTA DR1 data set. The periodogram power of a significant peak is 0.067 according to Monte Carlo simulations.
Hence there is no significant peak in this LP.

\subsection{CONSTRAINTS FROM PULSAR TIMING}
Since our analysis find no significant signal, we choose to set upper limits on the chirp masses of a few well-known SMBBH candidates.
A Monte Carlo analysis similar to that in \cite{jenet2004} was then used to place 95\% upper limits on the chirp mass of a few well-known SMBBH candidates. Before laying out our computational procedure, we wish to state that the 95\% upper limit on the chirp mass implies that the true value of the chirp mass is less than the upper limit with 95\% probability.
\begin{table*}
\begin{center}
\caption{Source information and upper limits. Column 1: name of the source; column 2: red shift; column 3: the observed orbital period; column 4: eccentricity; column 5: total mass; column 6: mass ratio; column 7: chirp mass; column 8: upper limit on chirp mass based on PPTA DR1 data set; column 9: upper limit on chirp mass based on PPTA P18 data set; column 10: ratio between $limit_{P18}$ and $M_{c}$; column 11: strain amplitude; column 12: evolution timescale; column 13: references. }
\begin{tabular}{cccccccccccccc}
\hline \hline
Name &$z$&$P$&$e$&$M_{total}$&$q$&$M_{c}$&$limit_{DR1}$&$limit_{P18}$&ratio&$h$&$\tau$&Ref.\\

 &  & [yr]  &    & [$10^8\,\rm M_{\odot}$]  &    &[$10^8\,\rm M_{\odot}$]&[$10^9\,\rm M_{\odot}$]&[$10^9\,\rm M_{\odot}$]& &[$10^{-17}$]&[Myr]&&\\
(1) & (2) & (3) & (4) & (5) & (6)  & (7) & (8) & (9) & (10) & (11) & (12) & (13) \\
\hline

NGC 5548 &0.0172  & 14.1  & 0.13   &2.7   &1.0     &1.2&7.6&2.4&20&6.4&33& 1&\\

Mrk 231  &0.0422  & 1.2   & -      & 1.5  &0.026   &0.16&15&6.0&370&0.45&1.4& 2&\\

OJ 287   &0.3056  & 12.1 & 0.700  &184   &0.008   &10.2&230&88&86&11.8&0.83& 3&\\

PG 1302-102  & 0.2784 & 5.2  & -      & 3.0  &-       &1.3&20&11&85&0.88&1.3 &4&\\

NGC 4151  &0.0033  & 15.9  & 0.42   & 0.56 &0.27    &0.19&14&2.0&110&1.4& 980&5&\\

Ark 120  &0.0327 &$\sim20$& -      & 2.6  &-       &1.1&43&15&140&2.2&100&6&\\ 

3C 66B   &0.0213  & 1.05 & -  &19   &0.58   &7.9&3.4&2.5&3.2&730&$1.4\times10^{-3}$& 7&\\
\hline \hline
\end{tabular}
\label{tab1}
\end{center}
Refs: (1) \citealt{2016ApJ...822....4L}, ApJ, 822, 4; (2) \citealt{2015ApJ...809..117Y}, ApJ, 809, 117; (3) \citealt{Valtonen2016}, ApJ,
819, L37; (4) \citealt{2015Natur.518...74G}, Nature, 518, 74; (5) \citealt{2012ApJ...759..118B}, ApJ, 759, 118; (6) \citealt{2019ApJS..241...33L}, ApJS, 241, 33;
(7) \citealt{Iguchi2010}, ApJ, 724, L166 
\end{table*}

The distance to PSR~J1909$-$3744 is $1.140\pm0.012$\,kpc \citep{2016MNRAS.455.1751R}. So the distribution of $d_p$ is taken as $(d_p/\rm{kpc})\in \mathcal{N}(1.140, 0.012)$.
The orbital frequency, $f$, and the SMBBH distance are calculated using values in Table \ref{tab1}.  
The eccentricity, $e$, uses the value in Table \ref{tab1} or taken as 0 when it is not available in electromagnetic observations (denoted as `-' in Table \ref{tab1}). Also the mass ratio, $q$, is taken as 1 when
it is not available from the electromagnetic observations.

The distribution for the other parameters are as follows:
$ l_0 \in U[0, 2 \pi],
\gamma \in U[0, 2 \pi], \iota \in U[0, \pi], \psi \in U[0, \pi]
$. 
For 3C 66B, we use the reported inclination angle in \cite{Iguchi2010},
i.e. $\iota \in U[0.08, 0.11]$ .
For NGC 5548, the full orbital parameters were reported with small error bars by \cite{2016ApJ...822....4L} and a schematic for the geometry of the SMBBH system can be found therein. 
The distribution for the parameters are as follows:
$(f/\rm{Hz}) \in U[2.15 \times 10^{-9}, 2.43 \times 10^{-9}], e \in U[0, 0.31], l_0 \in U[4.52, 5.22],
\gamma \in U[3.23, 3.84], \iota \in U[0.36, 0.48], \psi \in U[0, \pi]$.
In top panel of Figure~\ref {fig1}, we show 5 realizations of pre-fit timing residuals induced by NGC 5548 with the pulsar term included for each data set. For these 5 realizations, we use the upper limit chirp mass based on PPTA P18 data set, i.e. $2.4 \times 10^9\,\rm M_{\odot}$, to calculate timing residuals. The other parameters of each line are as follows:
\begin{enumerate}
\item red line: $(f/\rm{Hz}) = 2.40 \times 10^{-9}, e = 0.108, l_0 = 5.18, \gamma = 3.33, \iota = 0.421 , \psi = 0.58, d_p/\rm{kpc} = 1.1524
$
\item orange line: $(f/\rm{Hz}) = 2.36 \times 10^{-9}, e = 0.016, l_0 = 4.79, \gamma = 3.34, \iota = 0.364 , \psi = 3.08, d_p/\rm{kpc} = 1.1384
$
\item green line: $(f/\rm{Hz}) = 2.33 \times 10^{-9}, e = 0.231, l_0 = 5.21, \gamma = 3.64, \iota = 0.422 , \psi = 2.48, d_p/\rm{kpc} = 1.1453
$
\item blue line: $(f/\rm{Hz}) = 2.38 \times 10^{-9}, e = 0.0243, l_0 = 4.83, \gamma = 3.52, \iota = 0.455 , \psi = 0.89, d_p/\rm{kpc} = 1.1588
$
\item magenta line: $(f/\rm{Hz}) = 2.36 \times 10^{-9}, e = 0.218, l_0 = 4.96, \gamma = 3.76, \iota = 0.476 , \psi = 1.46, d_p/\rm{kpc} = 1.1405
$
\end{enumerate}
Timing residuals induced by other SMBBH candidates can be found in Appendix \ref{waveform}.

The corresponding waveform was generated using equations in Section~\ref{sec:signal}. 
The waveform was then added to the ToA data. 
Next we computed
the residuals using the TEMPO2
\citep{hobbs2006} package. 
The residuals were then analyzed using
the LP method described above. 
If a significant peak was found
(see above), then the signal was considered to be detected. A
total of 1000 waveforms were tested for each $M$. 
We adjust $M$ until $95\%$ of the injected signal can be detected.

In Table \ref{tab1}, we tabulate the upper limits on the chirp mass. For example, 95$\%$ confidence level upper limit on the chirp mass of NGC 5548 is $7.6 \times 10^9\,\rm M_{\odot}$ and $2.4 \times 10^9\,\rm M_{\odot}$ for PPTA DR1 and P18 data set respectively. And 95$\%$ confidence level upper limit on the chirp mass of 3C 66B is $2.5 \times 10^9\,\rm M_{\odot}$ for PPTA P18 data set. Our upper limits remain a factor of 3 to 370 above the reported values for these candidates.

\subsection{COMPARISON BETWEEN DR1 AND P18}

The DR1 data set for PSR~J1909$-$3744 has a time span of 5.5\,yrs with 125 data points and
P18 data set has a time span of 12.0\,yrs with 393 data points. Note that the time span of DR1 data set is even shorter than the 
period of the gravitational wave expected of NGC 5548, NGC 4151, OJ 287 and Ark 120, which reduce their detectabilities in such a data set significantly.

The corresponding upper
limit on the chirp mass for the P18 data set is typically a factor of 2 to 3 more stringent than that of DR1 data set.
The corresponding upper
limit on the strain amplitude for the P18 data set is typically a factor of 2 to 7 more stringent than that of DR1 data set.
 This is caused by two effects: (1) the P18 data set has more data points than DR1 data set, and
(2) the fitting process is known
to significantly reduce a PTA's sensitivity to GWs at the lowest
frequencies (because of the fit for pulsar spin period and its first time derivative).

We estimate the two effects respectively. According to the scaling relation~\ref{eqn:h} (see Appendix \ref{app} for derivation), 
\begin{equation}
\frac{h_{DR1}}{h_{P18}} \sim \sqrt{\frac{N_{P18}}{N_{DR1}}} =  \sqrt{\frac{393}{125}} \sim 2
\end{equation}
So the number of data points contributes about a factor of 2 and the fit for pulsar spin period and its first derivative contributes a factor of about 1 to 3. Specifically, the fitting process contributes a factor of about 3 for low frequency GW source NGC 5548 , OJ 287 and Ark 120. For NGC 4151, the DR1 upper limit is much worse, this is because low frequency combined with high eccentricity can reduce the sensitivity significantly. The fitting process can have a huge effect on GW detection of low frequency sources. A factor of 3 worse caused by the fitting process would require 9 times longer observation to compensate. Longer time span of PTA data is critical for detection of low frequency and eccentric SMBBHs.

\subsection{INFLUENCE OF EXTRA INFORMATION}
\label{sec:orbit_info}
\begin{figure}
\includegraphics[width=\textwidth]{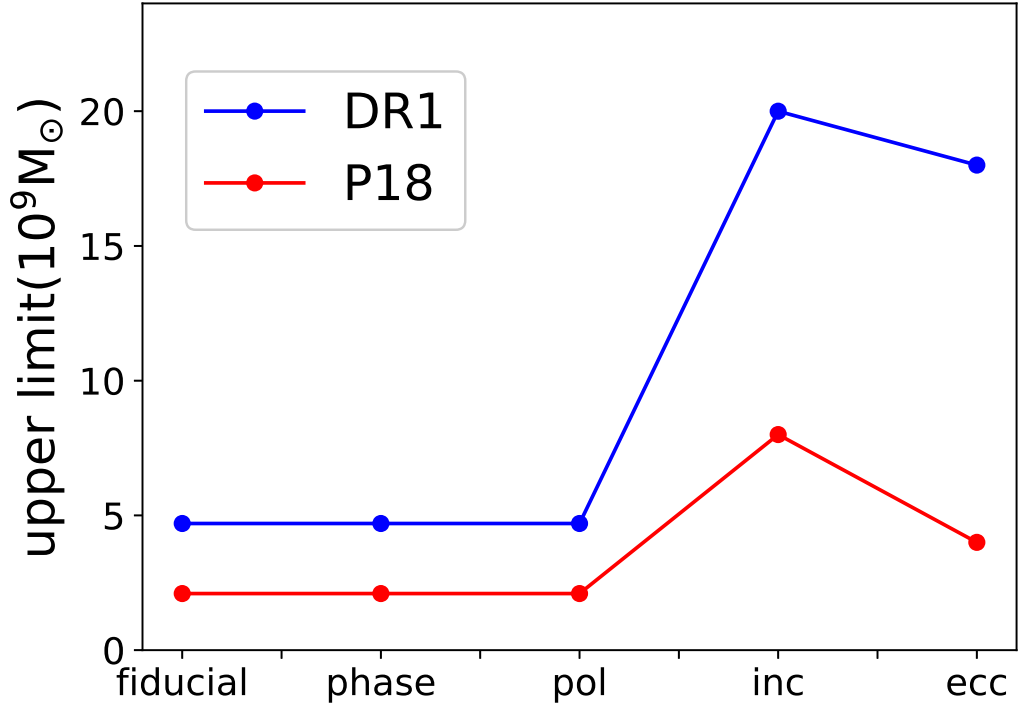}
\caption{Case study of how extra information about the orbital parameters of the SMBBH would influence the upper limits on the chirp mass. The blue dots are for DR1 data sets and the red dots are for P18 data sets. The horizontal axis labels 'fiducial', 'phase', 'pol', 'inc', 'ecc' represent fiducial case, phase effect, polarization angle effect, inclination effect, eccentricity effect respectively. See the text for details of the cases.}
\label{fig3}
\end{figure}

In this section, we discuss how extra information about the orbital parameters of the SMBBH would influence the upper limits on the strain amplitude. As we study the relative influence of the orbital parameters, so we just consider the Earth term of the signal.
We take NGC 5548 as an example and run the following cases for both DR1 and P18 data sets: 
\begin{enumerate}
\item fiducial case: We know everything about the GW source. $(f/\rm{Hz}) = 2.15 \times 10^{-9}, e = 0, l_0 = 0,
\gamma = 0, \iota = 0 , \psi = 0
$
\item phase effect: How varying phase influence the limit.
$(f/\rm{Hz}) = 2.15 \times 10^{-9}, e = 0, l_0 \in U[0, 2 \pi],
\gamma = 0, \iota = 0 , \psi = 0
$
\item polarization angle effect: How varying polarization angle influence the limit.
$(f/\rm{Hz}) = 2.15 \times 10^{-9}, e = 0, l_0 = 0,
\gamma = 0, \iota = 0 , \psi \in U[0, \pi] 
$
\item inclination effect: especially, how would edge on inclination influence the limit.
$(f/\rm{Hz}) = 2.15 \times 10^{-9}, e = 0, l_0 = 0,
\gamma = 0, \iota = 0.5 \pi , \psi \in U[0, \pi] 
$
\item eccentricity effect: how would large eccentricity influence the limit.
$(f/\rm{Hz}) = 2.15 \times 10^{-9}, e = 0.7, l_0 = 0,
\gamma = 0, \iota = 0 , \psi \in U[0, \pi] 
$
\end{enumerate}

The results are shown in Figure \ref{fig3}.
Comparing case 1 and case 2, different phases do not influence upper limits.
For case 3, polarization angle effect is equivalent to phase effect when $\iota = 0$, so the result of case 3 is the same as case 2. 
Comparing case 3 and 4, different inclinations can cause upper limits on the strain amplitude to vary by a factor about 10. We get a worse upper limit when $\cos \iota \sim 0$.
Comparing case 3 and 5, increasing the eccentricity leads to a worse upper limit. This is because the LS power is less when the eccentricity gets larger.

\subsection{INFLUENCE OF THE SOLAR SYSTEM EPHEMERIS}
\label{sec:sse}
\begin{figure}
\includegraphics[width=\textwidth]{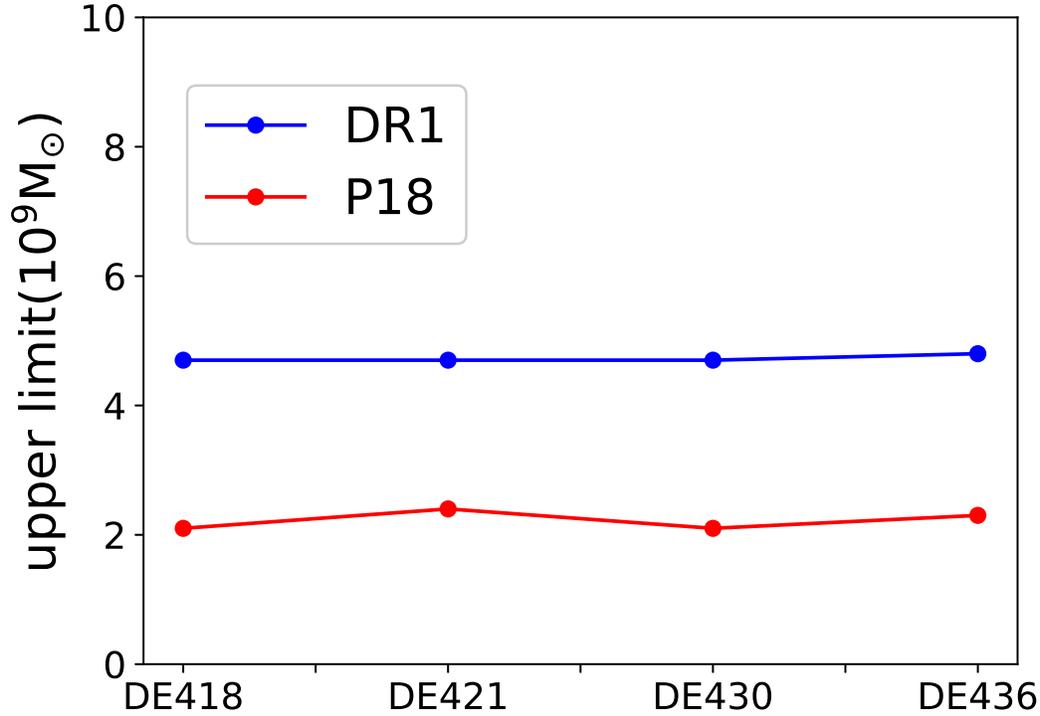}
\caption{Upper limits on the chirp mass of NGC 5548 using different Solar System ephemeris. The blue dots are for DR1 data sets and the red dots are for P18 data sets.}
\label{fig4}
\end{figure}

SSE errors can mimic a GWB signal \citep{2016MNRAS.455.4339T}. Recently, it was shown that the GWB constraints are sensitive to the SSE model used \citep{2018ApJ...859...47A}. The GWB constraints are very different if they use different SSE models. If they take the SSE errors into account, the constraints would converge. However, this removes most evidence for the presence of a GWB. 

To investigate how SSE model used would influence the upper limits on the strain amplitude, we repeated the fiducial case in Section~\ref{sec:orbit_info} using each of the JPL DE418, JPL DE421, JPL DE430 and JPL DE436 SSEs for both DR1 and P18 data sets. The results are shown in Figure \ref{fig4}. It indicates that the choice of SSE does not influence the upper limits on the chirp mass of NGC 5548. This is consistent with our expectation due to the intrinsic difference between the two detection experiment. The GWB ‘signal’ is a kind of noise in the PTA data sets, thus its detection is more prone to inaccurate or incomplete characterization of various sources of noises, including that of SSE.

\section{CONCLUSION}
\label{sec:discussion}
We have performed a search for GWs from several SMBBH candidates in long-term timing observations of the pulsar PSR~J1909$-$3744 as part of the  PPTA. No statistically significant signals were found. We constrained the chirp masses for these SMBBH candidates. For example, the 95$\%$ confidence level upper limit on the chirp mass of NGC 5548 is $2.4 \times 10^9\,\rm M_{\odot}$. The corresponding limit for 3C 66B is $2.5 \times 10^9\,\rm M_{\odot}$. Our upper limits remain a factor of 3 $\sim 370$ above the reported values for these candidates.
The corresponding upper limit on the strain amplitude for PPTA P18 data set is about a factor of 2 to 7 more stringent than that of DR1 data set. The number of data points contributes about a factor of 2 and the fit for pulsar spin period and its first derivative contributes about a factor of 1 to 3. Longer time span of PTA data is critical for detection of low frequency and eccentric SMBBHs.
We investigated how information about the orbital parameters would influence the upper limit.
This may suggest strategies for future PTA single GW source searches. If a PTA's goal is to detect a known 
single GW source, we should consider the full orbital 
parameters to decide which source as our detection target.

Although it was shown that the GWB constraints are sensitive to the SSE model used \citep{2018ApJ...859...47A} very recently, we find that our upper limits on the chirp mass of NGC 5548 is not sensitive to the SSE model used. But whether this result only holds for LP method or single pulsar analysis is not known. We do not attempt to answer this question in this paper but leave it to a future study.

We note that there are ongoing large time-domain surveys that have yielded hundreds of quasar and AGN
candidates with possible periodic variability.
The list of best candidates is growing.
Future work will place constraints on other known nearby and new candidate SMBBH systems.

In this work, we just used one pulsar, PSR~J1909$-$3744, which has the second best timing precision in the PPTA data set. PSR~J0437$-$4715 has the best timing precision, but the timing residuals of PSR~J0437$-$4715 has a strong red noise component and are not suitable for the LP method. \cite{2015Sci...349.1522S} found no evidence for red noise for PSR~J1909$-$3744, so PSR~J1909$-$3744 is suitable for the LP method. Current PTA's sensitivity to GWs is often dominated by several best-timing-precision pulsars. 
For example, \cite{2016MNRAS.455.1665B} pointed out that the array is heavily dominated by PSR J1909-3744,
contributing more than 60 percent of $(\rm S/N)^2$, followed by PSR J1713+0747 at about 20 percent.
Moreover, \cite{2015Sci...349.1522S} found that their limit placed using PSR~J1909$-$3744 is the same as the joint limit, indicating that PSR~J1909$-$3744 is the dominant pulsar in the sample. Although the bounding techniques for single GW and GWB are different, PSR~J1909$-$3744 contributes about 30~\% of the total S/N of the full PPTA data set and the upper chirp mass limits given by the one pulsar analysis are meaningful even compared with a full PPTA analysis. However, if the opening angle between the pulsar and the SMBBH is large, it is not sensitive to the SMBBH. In this case, the constraints can be bad, and the full PPTA data set with a better sky coverage can be used to place better constraints. Nevertheless, the one pulsar analysis provides valuable insights into problems such as the influence of the SSE, etc. We defer analysis using the full PPTA data set to future work. We will also discuss how adding pulsars to present PTAs would influence the constraints.

\normalem
\begin{acknowledgements}
We thank the anonymous referee for very useful comments on the manuscript. This work is supported by National Key R$\&$D Program of China No. 2017YFA0402600, by the CAS International Partnership Program No.114-
A11KYSB20160008, by the CAS Strategic Priority Research Program No. XDB23000000, and by NSFC grant No. 11725313 and 11690024.
Y.R.L. acknowledges financial support from the NSFC (11573026) and the Youth Innovation Promotion Association CAS. J.M.W acknowledges financial support from the NSFC (11833008), the National Key R\&D Program of China (2016YFA0400701), and the Key Research Program of Frontier Sciences of the Chinese Academy of Sciences (QYZDJ-SSWSLH007). The Parkes radio telescope is part of the Australia Telescope, which is funded by the Commonwealth of Australia for operation as a National Facility managed by the Commonwealth Scientific and Industrial Research Organisation (CSIRO). This work is supported by Aliyun Cloud.
The authors wish to acknowledge crucial guidance from George Hobbs and many useful discussions with Xingjiang Zhu, Zhoujian Cao and Youjun Lu.    
\end{acknowledgements}

\appendix
\section{scaling relation}
\label{app}
$\rm SNR \geq 5$ is usually required for a definitive detection, i.e.
\begin{equation}
\sqrt{(s|s)} \geq 5.
\end{equation}
This formular can be approximated as 
\begin{equation}
\sqrt{\frac{1}{2}N\frac{A^2}{\sigma^2}} \geq 5,
\end{equation}
where $N$ is the total number of data points, $A$ is the amplitude of the timing residuals induced by the SMBBH, $\sigma$ is the noise rms.
The thresold of the amplitude to be detected is 
\begin{equation}
A_{th} = \sqrt{\frac{50}{N}}\sigma,
\end{equation}
i.e. if $A$ is larger than $A_{th}$, then we can detect the signal.
We can detect the signal at 95 confidence level when $A_{0.05}= A_{th}$, where $A_{0.05}$ is 0.05 fractile of $A$ .
95\% upper limit of M($M_{95}$) and h($h_{95}$) is determined by $A_{0.05}$ through the following formular
\begin{equation}
f_M(M_{95}) = f_h(h_{95}) = A_{0.05} = A_{th}
\end{equation}
So the functions $f_M$ and $f_h$ are essential to determine $M_{95}$ and $h_{95}$. 
Assuming the fiducial parameters:
$e = 0, l_0 = 0, \gamma_0 = 0$,
the Earth term pulsar timing residuals can be written as
\begin{equation}
\begin{split}
s = &\epsilon(1+\cos^2\iota)(F^{+}\cos2 \psi + F^{\times}\sin2 \psi)\sin(nl(t)) \\
&+ \epsilon (2\cos\iota)(F^{+}\sin2 \psi -F^{\times}\cos2 \psi)\cos(nl(t)),
\end{split}
\end{equation}
where
\begin{equation}
\epsilon = \frac{(GM)^{5/3}}{c^4D_L(2\pi f)^{1/3}}.
\end{equation}
The strain amplitude is given by
\begin{equation}
h = \frac{2(GM)^{5/3}(2\pi f)^{2/3}}{c^4D_L},
\end{equation}
so 
\begin{equation}
h_{95} \propto \epsilon_{95} \propto M_{95}^{5/3}. 
\end{equation}
The amplitude $A$ is calculated by
\begin{equation}
\begin{split}
A^2 = \lVert s \rVert^2 = &\epsilon^2[(1+\cos^2\iota)^2(F^{+}\cos2 \psi + F^{\times}\sin2 \psi)^2\\
&+(2\cos\iota)^2(F^{+}\sin2 \psi -F^{\times}\cos2 \psi)^2].
\end{split}
\end{equation}
Further assuming $\iota = 0$, the amplitude can be simplified as:
\begin{equation}
A = 2\epsilon\sqrt{F^{+2}+F^{\times2}}=(1+\cos\theta)\epsilon.
\end{equation}
Combining the relations above, we have
\begin{equation}
\label{eqn:h}
h_{95} \propto \epsilon_{95} \propto A_{th} \propto \frac{1}{\sqrt{N}}
\end{equation}
 
\clearpage  
\section{Gw-induced timing residuals}
\label{waveform}
\begin{figure*}[h]
\includegraphics[width=1\textwidth]{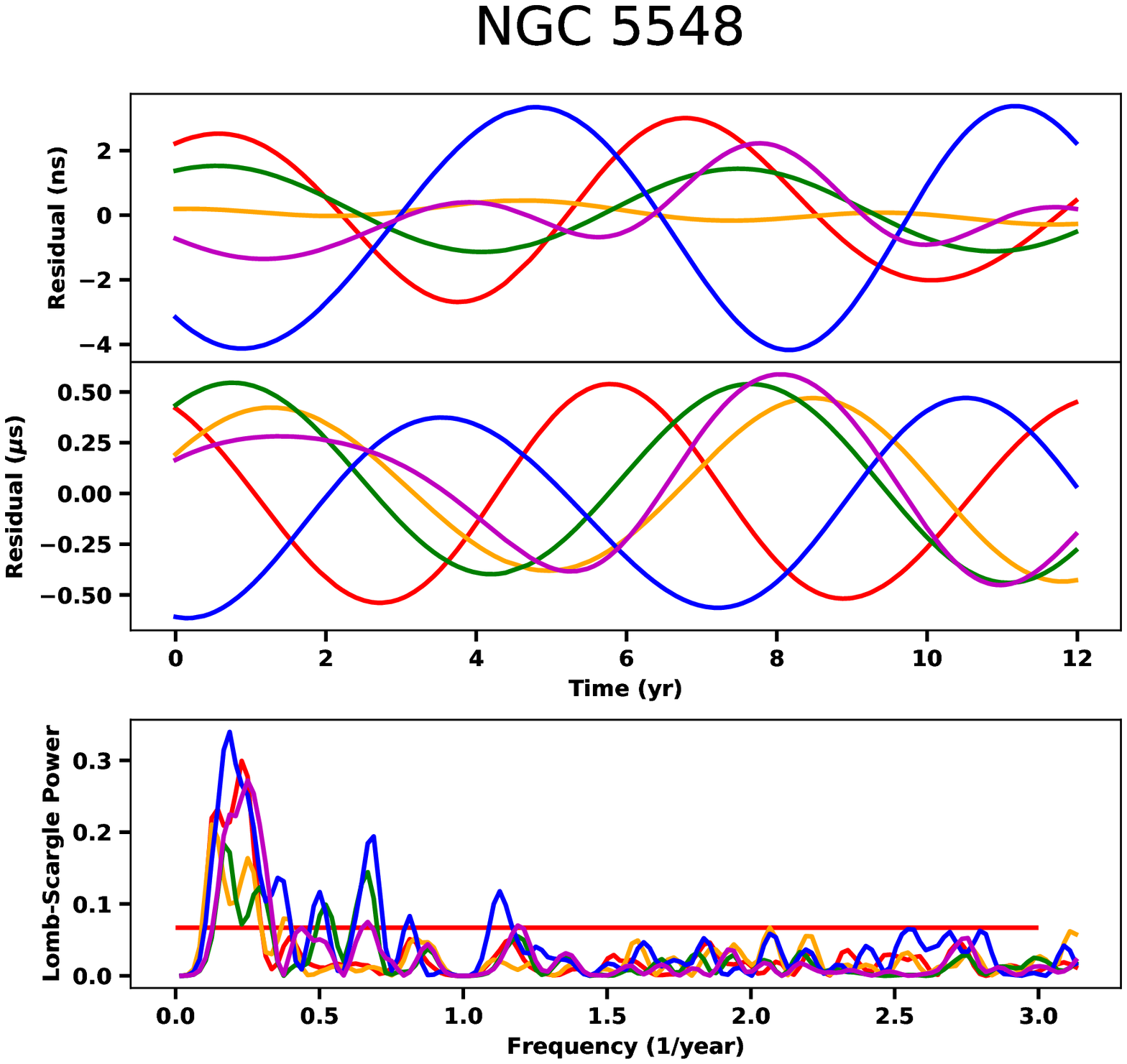}
\caption{Top: 5 realizations of pre-fit timing residuals induced by NGC 5548 with the pulsar term included using the chirp mass estimated from other approaches. Middle: 5 realizations of pre-fit timing residuals induced by NGC 5548 with the pulsar term included using the upper limit chirp mass based on PPTA P18 data set. 
Bottom: The corresponding Lomb-Scargle periodogram for each realization in the middle panel. The horizontal red line corresponds to our detection threshold of $\rm{FAP} = 10^{-3}$.
}
\label{ngc5548}
\end{figure*}
The parameters of each line are as follows:
\begin{enumerate}
\item red line: $(f/\rm{Hz}) = 2.42 \times 10^{-9}, e = 0.061, l_0 = 4.66, \gamma = 3.49, \iota = 0.361 , \psi = 1.64, d_p/\rm{kpc} = 1.1322
$
\item orange line: $(f/\rm{Hz}) = 2.25 \times 10^{-9}, e = 0.035, l_0 = 4.97, \gamma = 3.48, \iota = 0.389 , \psi = 1.06, d_p/\rm{kpc} = 1.1330
$
\item green line: $(f/\rm{Hz}) = 2.28 \times 10^{-9}, e = 0.023, l_0 = 5.10, \gamma = 3.78, \iota = 0.450 , \psi = 0.76, d_p/\rm{kpc} = 1.1382
$
\item blue line: $(f/\rm{Hz}) = 2.30 \times 10^{-9}, e = 0.102, l_0 = 4.80, \gamma = 3.45, \iota = 0.447 , \psi = 3.89, d_p/\rm{kpc} = 1.1562
$
\item magenta line: $(f/\rm{Hz}) = 2.41 \times 10^{-9}, e = 0.216, l_0 = 5.14, \gamma = 3.55, \iota = 0.431 , \psi = 2.72, d_p/\rm{kpc} = 1.1481
$
\end{enumerate}
\clearpage

\begin{figure*}[h]
\includegraphics[width=1\textwidth]{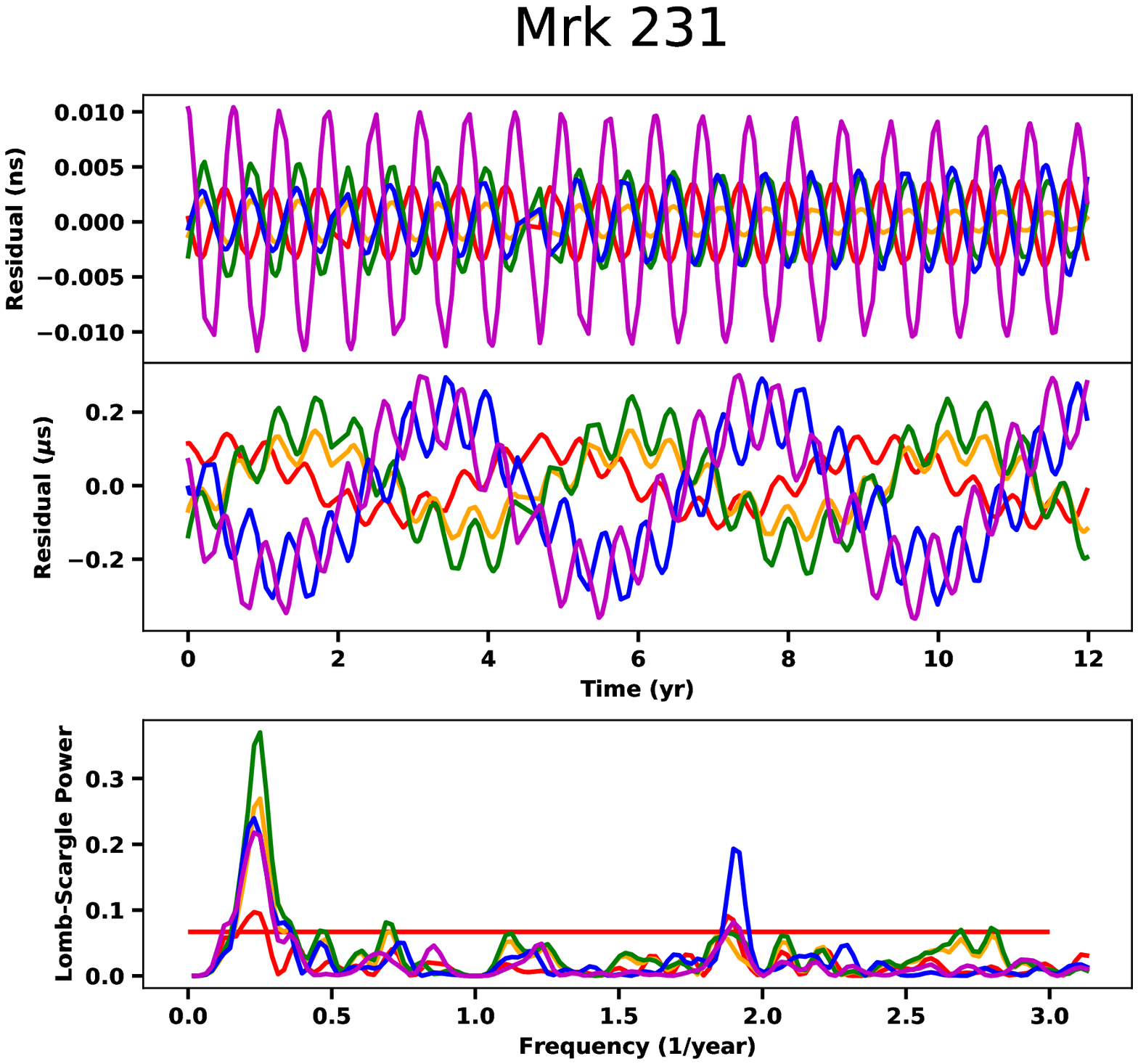}
\caption{Top: 5 realizations of pre-fit timing residuals induced by Mrk 231 with the pulsar term included using the chirp mass estimated from other approaches. Middle: 5 realizations of pre-fit timing residuals induced by Mrk 231 with the pulsar term included using the upper limit chirp mass based on PPTA P18 data set. 
Bottom: The corresponding Lomb-Scargle periodogram for each realization in the middle panel. The horizontal red line corresponds to our detection threshold of $\rm{FAP} = 10^{-3}$.
}
\label{mrk231}
\end{figure*}
The parameters of each line are as follows:
\begin{enumerate}
\item red line: $(f/\rm{Hz}) = 2.53 \times 10^{-8}, e = 0, l_0 = 3.39, \gamma = 5.98, \iota = 1.22 , \psi = 0.60, d_p/\rm{kpc} = 1.1348
$
\item orange line: $(f/\rm{Hz}) = 2.53 \times 10^{-8}, e = 0, l_0 = 4.33, \gamma = 0.67, \iota = 1.41 , \psi = 0.93, d_p/\rm{kpc} = 1.1469
$
\item green line: $(f/\rm{Hz}) = 2.53 \times 10^{-8}, e = 0, l_0 = 0.34, \gamma = 3.87, \iota = 0.78 , \psi = 1.80, d_p/\rm{kpc} = 1.1460
$
\item blue line: $(f/\rm{Hz}) = 2.53 \times 10^{-8}, e = 0, l_0 = 5.02, \gamma = 3.94, \iota = 2.73 , \psi = 0.06, d_p/\rm{kpc} = 1.1332
$
\item magenta line: $(f/\rm{Hz}) = 2.53 \times 10^{-8}, e = 0, l_0 = 1.52, \gamma = 3.48, \iota = 0.11 , \psi = 2.00, d_p/\rm{kpc} = 1.1208
$
\end{enumerate}
\clearpage

\begin{figure*}[h]
\includegraphics[width=1\textwidth]{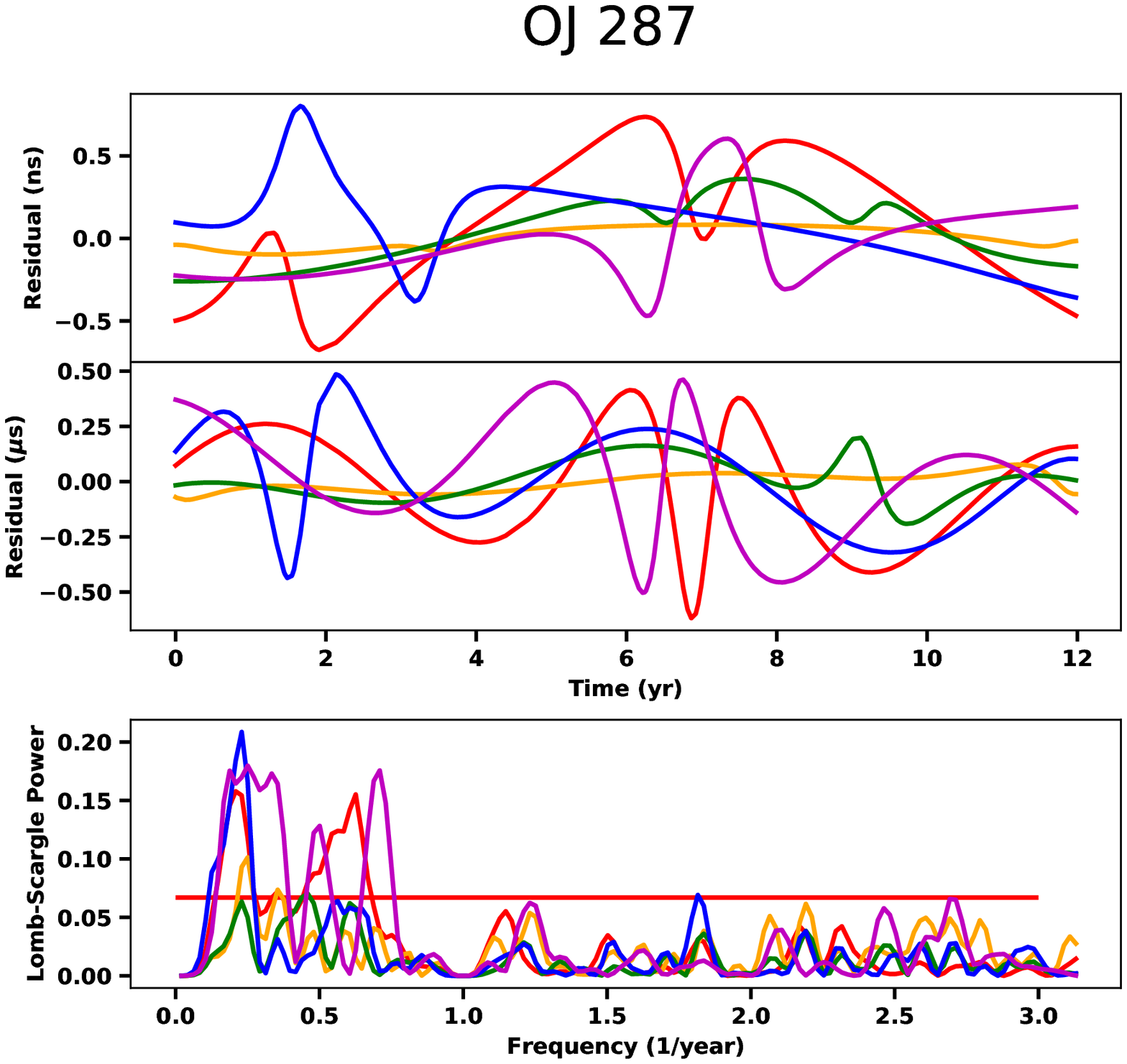}
\caption{Top: 5 realizations of pre-fit timing residuals induced by OJ 287 with the pulsar term included using the chirp mass estimated from other approaches. Middle: 5 realizations of pre-fit timing residuals induced by OJ 287 with the pulsar term included using the upper limit chirp mass based on PPTA P18 data set. 
Bottom: The corresponding Lomb-Scargle periodogram for each realization in the middle panel. The horizontal red line corresponds to our detection threshold of $\rm{FAP} = 10^{-3}$.
}
\label{oj287}
\end{figure*}
The parameters of each line are as follows:
\begin{enumerate}
\item red line: $(f/\rm{Hz}) = 2.62 \times 10^{-9}, e = 0.7, l_0 = 2.85, \gamma = 0.11, \iota = 0.29 , \psi = 2.33, d_p/\rm{kpc} = 1.1378
$
\item orange line: $(f/\rm{Hz}) = 2.62 \times 10^{-9}, e = 0.7, l_0 = 0.27, \gamma = 4.60, \iota = 1.54 , \psi = 1.68, d_p/\rm{kpc} = 1.1526
$
\item green line: $(f/\rm{Hz}) = 2.62 \times 10^{-9}, e = 0.7, l_0 = 1.67, \gamma = 4.57, \iota = 1.79 , \psi = 2.96, d_p/\rm{kpc} = 1.1408
$
\item blue line: $(f/\rm{Hz}) = 2.62 \times 10^{-9}, e = 0.7, l_0 = 5.62, \gamma = 0.38, \iota = 0.70 , \psi = 0.73, d_p/\rm{kpc} = 1.1522
$
\item magenta line: $(f/\rm{Hz}) = 2.62 \times 10^{-9}, e = 0.7, l_0 = 3.10, \gamma = 3.41, \iota = 2.79 , \psi = 1.91, d_p/\rm{kpc} = 1.1493
$
\end{enumerate}
\clearpage

\begin{figure*}[h]
\includegraphics[width=1\textwidth]{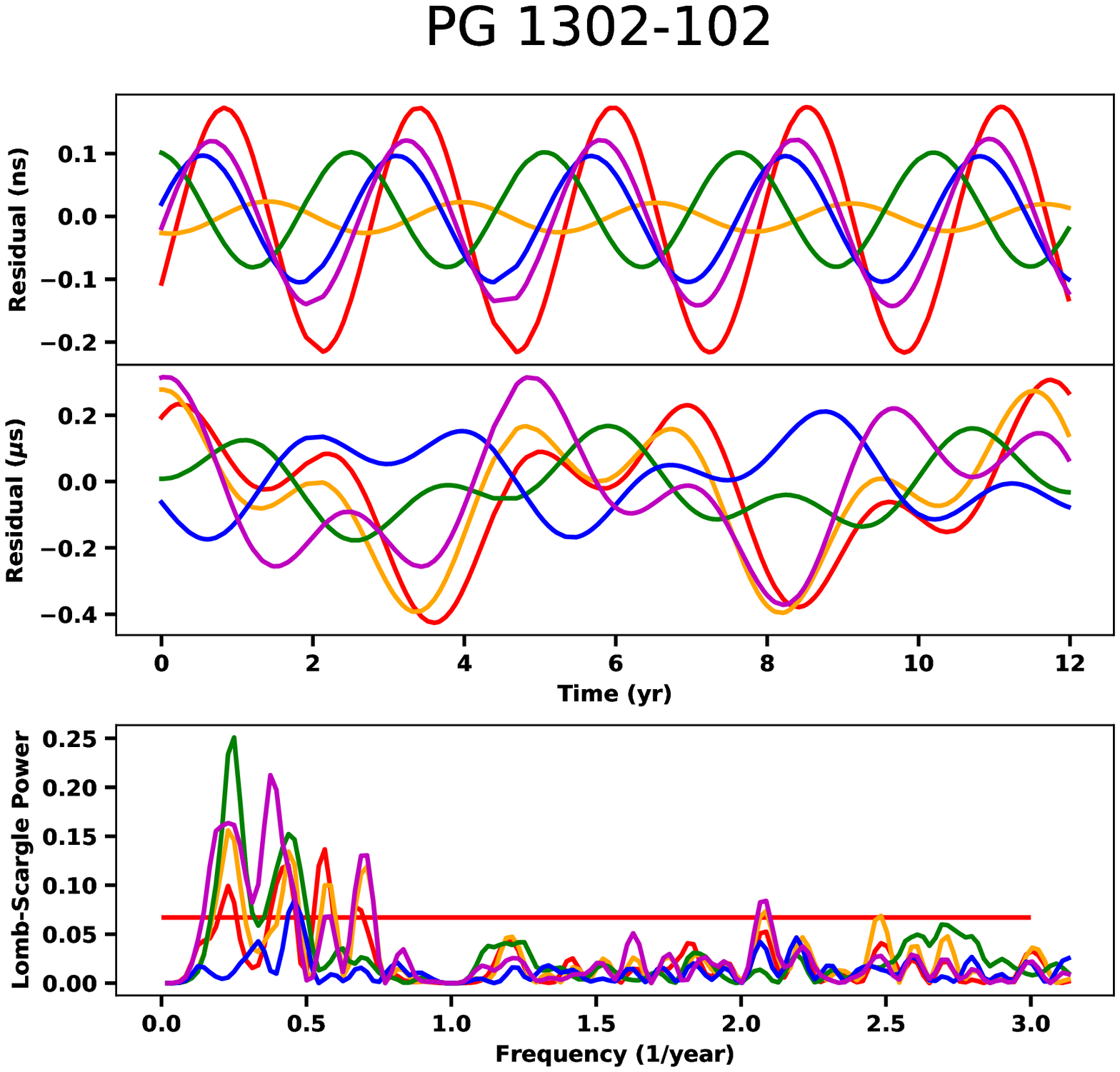}
\caption{Top: 5 realizations of pre-fit timing residuals induced by PG 1302-102 with the pulsar term included using the chirp mass estimated from other approaches. Middle: 5 realizations of pre-fit timing residuals induced by PG 1302-102 with the pulsar term included using the upper limit chirp mass based on PPTA P18 data set.
Bottom: The corresponding Lomb-Scargle periodogram for each realization in the middle panel. The horizontal red line corresponds to our detection threshold of $\rm{FAP} = 10^{-3}$.
}
\label{pg1302-102}
\end{figure*}
The parameters of each line are as follows:
\begin{enumerate}
\item red line: $(f/\rm{Hz}) = 6.15 \times 10^{-9}, e = 0, l_0 = 2.93, \gamma = 3.07, \iota = 0.14 , \psi = 0.81, d_p/\rm{kpc} = 1.1332
$
\item orange line: $(f/\rm{Hz}) = 6.15 \times 10^{-9}, e = 0, l_0 = 4.92, \gamma = 2.62, \iota = 2.83 , \psi = 2.35, d_p/\rm{kpc} = 1.1477
$
\item green line: $(f/\rm{Hz}) = 6.15 \times 10^{-9}, e = 0, l_0 = 1.11, \gamma = 5.12, \iota = 2.00 , \psi = 2.91, d_p/\rm{kpc} = 1.1448
$
\item blue line: $(f/\rm{Hz}) = 6.15 \times 10^{-9}, e = 0, l_0 = 6.14, \gamma = 2.19, \iota = 1.89 , \psi = 2.59, d_p/\rm{kpc} = 1.13926
$
\item magenta line: $(f/\rm{Hz}) = 6.15 \times 10^{-9}, e = 0, l_0 = 5.83, \gamma = 6.09, \iota = 0.21 , \psi = 1.13, d_p/\rm{kpc} = 1.1401
$
\end{enumerate}
\clearpage

\begin{figure*}[h]
\includegraphics[width=1\textwidth]{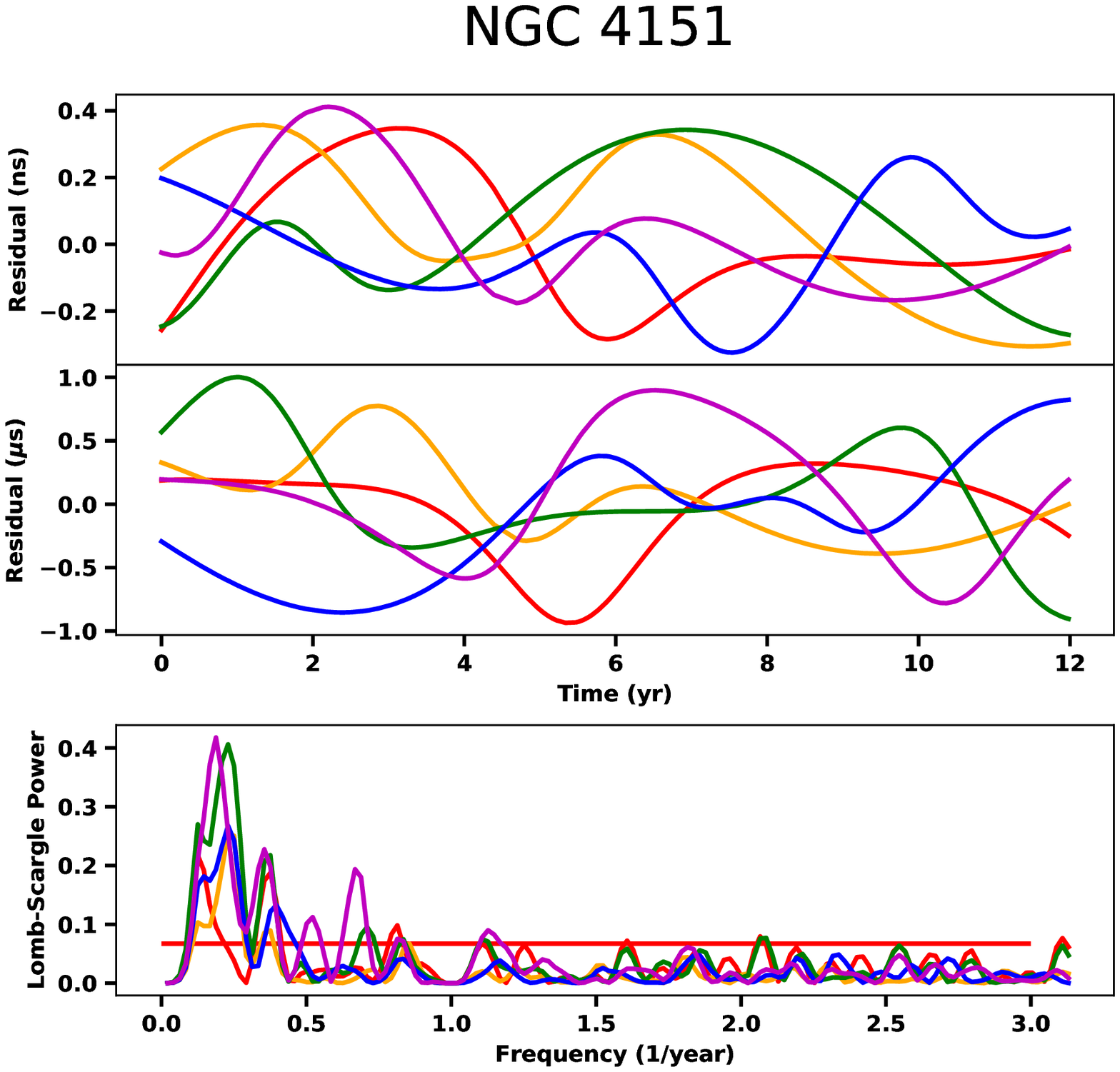}
\caption{Top: 5 realizations of pre-fit timing residuals induced by NGC 4151 with the pulsar term included using the chirp mass estimated from other approaches. Middle: 5 realizations of pre-fit timing residuals induced by NGC 4151 with the pulsar term included using the upper limit chirp mass based on PPTA P18 data set.
Bottom: The corresponding Lomb-Scargle periodogram for each realization in the middle panel. The horizontal red line corresponds to our detection threshold of $\rm{FAP} = 10^{-3}$.
}
\label{ngc4151}
\end{figure*}
The parameters of each line are as follows:
\begin{enumerate}
\item red line: $(f/\rm{Hz}) = 2.01 \times 10^{-9}, e = 0.42, l_0 = 2.39, \gamma = 0.55, \iota = 0.63 , \psi = 2.68, d_p/\rm{kpc} = 1.1243
$
\item orange line: $(f/\rm{Hz}) = 2.01 \times 10^{-9}, e = 0.42, l_0 = 2.54, \gamma = 3.57, \iota = 0.66 , \psi = 0.09, d_p/\rm{kpc} = 1.1252
$
\item green line: $(f/\rm{Hz}) = 2.01 \times 10^{-9}, e = 0.42, l_0 = 3.87, \gamma = 5.82, \iota = 2.85 , \psi = 2.72, d_p/\rm{kpc} = 1.1406
$
\item blue line: $(f/\rm{Hz}) = 2.01 \times 10^{-9}, e = 0.42, l_0 = 1.96, \gamma = 2.36, \iota = 2.58 , \psi = 2.19, d_p/\rm{kpc} = 1.1263
$
\item magenta line: $(f/\rm{Hz}) = 2.01 \times 10^{-9}, e = 0.42, l_0 = 2.62, \gamma = 5.63, \iota = 0.54 , \psi = 1.29, d_p/\rm{kpc} = 1.1497
$
\end{enumerate}
\clearpage

\begin{figure*}[h]
\includegraphics[width=1\textwidth]{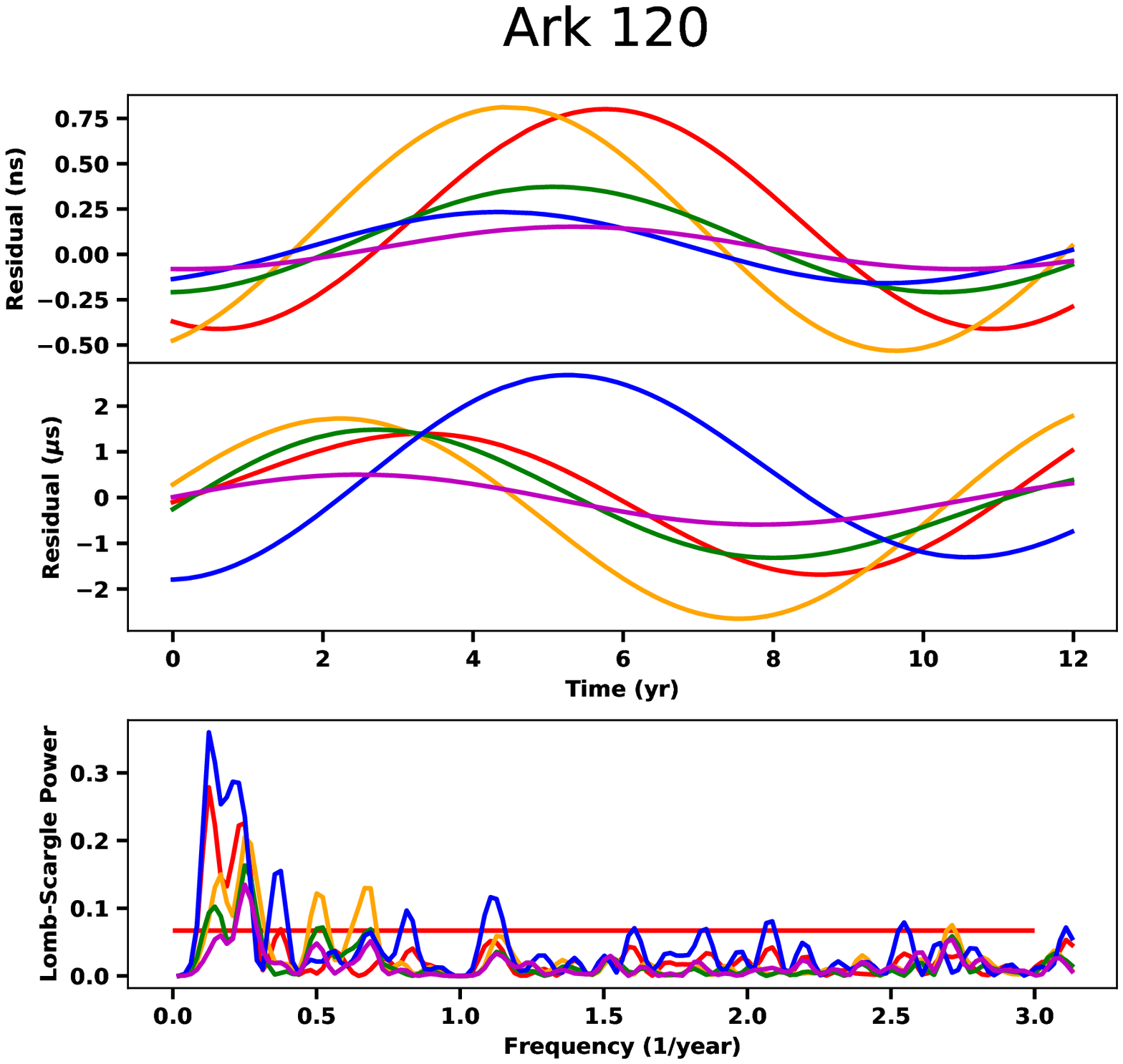}
\caption{Top: 5 realizations of pre-fit timing residuals induced by Ark 120 with the pulsar term included using the chirp mass estimated from other approaches. Middle: 5 realizations of pre-fit timing residuals induced by Ark 120 with the pulsar term included using the upper limit chirp mass based on PPTA P18 data set.
Bottom: The corresponding Lomb-Scargle periodogram for each realization in the middle panel. The horizontal red line corresponds to our detection threshold of $\rm{FAP} = 10^{-3}$.
}
\label{ark120}
\end{figure*}
The parameters of each line are as follows:
\begin{enumerate}
\item red line: $(f/\rm{Hz}) = 1.53 \times 10^{-9}, e = 0, l_0 = 5.75, \gamma = 1.78, \iota = 3.08 , \psi = 0.26, d_p/\rm{kpc} = 1.1425
$
\item orange line: $(f/\rm{Hz}) = 1.53 \times 10^{-9}, e = 0, l_0 = 5.06, \gamma = 3.44, \iota = 0.10 , \psi = 1.34, d_p/\rm{kpc} = 1.1259
$
\item green line: $(f/\rm{Hz}) = 1.53 \times 10^{-9}, e = 0, l_0 = 3.71, \gamma = 5.47, \iota = 0.60 , \psi = 1.10, d_p/\rm{kpc} = 1.1415
$
\item blue line: $(f/\rm{Hz}) = 1.53 \times 10^{-9}, e = 0, l_0 = 1.58, \gamma = 1.44, \iota = 0.29 , \psi = 0.04, d_p/\rm{kpc} = 1.1476
$
\item magenta line: $(f/\rm{Hz}) = 1.53 \times 10^{-9}, e = 0, l_0 = 2.86, \gamma = 5.37, \iota = 1.34 , \psi = 1.80, d_p/\rm{kpc} = 1.1432
$
\end{enumerate}
\clearpage

\begin{figure*}[h]
\includegraphics[width=1\textwidth]{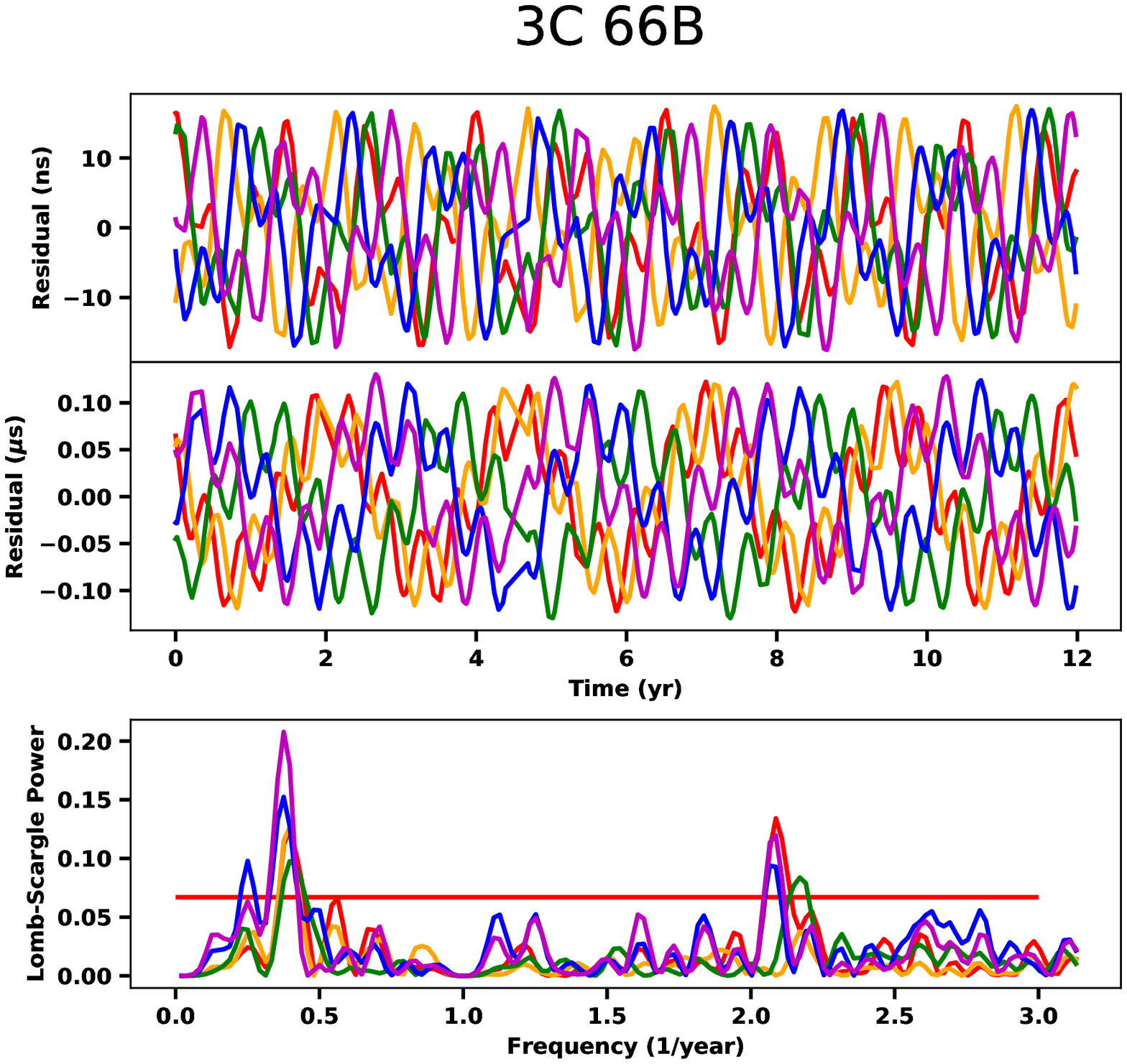}
\caption{Top: 5 realizations of pre-fit timing residuals induced by 3C 66B with the pulsar term included using the chirp mass estimated from other approaches. Middle: 5 realizations of pre-fit timing residuals induced by 3C 66B with the pulsar term included using the upper limit chirp mass based on PPTA P18 data set.
Bottom: The corresponding Lomb-Scargle periodogram for each realization in the middle panel. The horizontal red line corresponds to our detection threshold of $\rm{FAP} = 10^{-3}$.
}
\label{3c66b}
\end{figure*}
The parameters of each line are as follows:
\begin{enumerate}
\item red line: $(f/\rm{Hz}) = 3.02 \times 10^{-8}, e = 0, l_0 = 5.43, \gamma = 2.03, \iota = 0.11 , \psi = 0.01, d_p/\rm{kpc} = 1.1276
$
\item orange line: $(f/\rm{Hz}) = 3.02 \times 10^{-8}, e = 0, l_0 = 1.06, \gamma = 3.95, \iota = 0.09 , \psi = 1.50, d_p/\rm{kpc} = 1.1428
$
\item green line: $(f/\rm{Hz}) = 3.02 \times 10^{-8}, e = 0, l_0 = 3.41, \gamma = 6.00, \iota = 0.11 , \psi = 0.67, d_p/\rm{kpc} = 1.1432
$
\item blue line: $(f/\rm{Hz}) = 3.02 \times 10^{-8}, e = 0, l_0 = 2.05, \gamma = 1.08, \iota = 0.10 , \psi = 2.13, d_p/\rm{kpc} = 1.1660
$
\item magenta line: $(f/\rm{Hz}) = 3.02 \times 10^{-8}, e = 0, l_0 = 0.21, \gamma = 3.87, \iota = 0.08 , \psi = 1.05, d_p/\rm{kpc} = 1.1316
$
\end{enumerate}
\clearpage
\bibliographystyle{raa}
\bibliography{cgw}

\end{document}